\documentclass[]{aastex631}

\shorttitle{Model-independent classification of events from the first CHIME/FRB Fast Radio Burst catalog}
\shortauthors{A. Chaikova, D. Kostunin and S. Popov}
\graphicspath{{./}{figures/}}

\begin{document}

\title{Model-independent classification of events from the first CHIME/FRB Fast Radio Burst catalog}

\correspondingauthor{Dmitriy Kostunin}
\email{dmitriy.kostunin@desy.de}

\author[ 0000-0002-7902-6179 ]{Anastasia Chaikova}
\affiliation{HSE University - St. Petersburg, 194100 St. Petersburg, Russia}
\affiliation{JetBrains Research, 194100 St. Petersburg, Russia}

\author[0000-0002-0487-0076]{Dmitriy Kostunin}
\affiliation{JetBrains Research, 194100 St. Petersburg, Russia}
\affiliation{DESY, 15738 Zeuthen, Germany}

\author[0000-0002-4292-8638]{Sergei B. Popov}
\affiliation{Lomonosov Moscow State University, Sternberg Astronomical Institute. Universitetski pr. 13, 119234 Moscow, Russia}



\begin{abstract}
The CHIME/FRB collaboration has recently published a catalog containing about half a thousand fast radio bursts (FRBs) including their spectra and several reconstructed properties, like signal widths, amplitudes, etc. 
We have developed a model-independent approach for a classification of these bursts using cross-correlation and clustering algorithms applied to one-dimensional intensity profiles of the bursts (i.e., to amplitudes as a function of time averaged over the frequency). 
Using this algorithm we identified two major classes of FRBs featuring different waveform morphology, and, simultaneously, different distribution of brightness temperature. 
Bursts from one of the identified cluster have lower brightness temperatures and larger widths than events from the other cluster. Both groups include bursts from the repeaters and one-off events.
\end{abstract}



\section{Introduction} \label{sec:intro}

Fast radio bursts (FRBs) are millisecond-scale radio flares of high intensity (see a recent review, e.g., in \citealt{2021Univ....7...76N}).  
Thousands of such events have been detected to date. 
Among them, there are hundreds of individual flares which were detected from some sources only once and numerous events from sources producing repeating bursts. 
In few cases, monitoring of prolific repeaters with high sensitivity instruments, like FAST, allowed detection of many hundreds of FRBs from a given source (e.g., \citealt{2021arXiv211111764X}). 

One of the best instruments for FRB studies is the CHIME telescope operating in a special FRB mode~\citep{2018ApJ...863...48C}.
Recently, the first catalog of FRBs detected with this radio telescope was made public \citep{CHIMEFRB:2021srp}. 
The version of the catalog that we use contains information about 536 events, among which 62 are bursts from the repeaters. 
Already within this relatively small sample of  FRBs one can see that the bursts are very different in their morphology, which is discussed already, e.g.,~\cite{2021ApJ...923....1P}.
The main idea of our data analysis is to find visible and hidden (because of the low signal-to-noise ratio) model-independent similarities between the events by correlating their waveforms and grouping (clustering) these events using a cross-correlation as a measure of distance between the groups (clusters).
In this research note we briefly describe the approach, present our first results, and discuss them.

\section{Data processing and clustering} \label{sec:dataproc}
In the present work we use a simplified approach implying analysis of the waveforms derived from the data provided by the catalog.
The original dataset contains calibrated ``waterfall'' data, i.e., spectral densities of 16384 frequency channels (400 to 800 MHz) as a function of time $A(f,t)$ with sampling of 0.983\,ms.
We average these arrays over the frequency and obtain a one-dimensional array: $A(f,t) \to A(t) = \langle A(f,t)\rangle_f$.
This averaging helps us to keep the relative amplitudes on a right level since some frequency bands are zeroed to remove radio frequency interference (RFI).

Since we process the discrete data, we define a cross-correlation $C(A,B)$ between waveforms $A[i]$ and $B[j]$ represented by arrays as a maximum of discrete convolution: ${C(A,B) = \max(A \; * \; B) = \max \sum_m A[m]B[m+n]}$.
To improve the sensitivity of the cross-correlation, the actual sequence includes the following preprocessing steps:
\begin{enumerate}
\item subtracting the baseline $A \to A - \langle A \rangle$ decreases the contribution of the noise to the total cross-correlation;
\item normalizing the waveforms $A \to A \times (\sum_i A[i]^2)^{-1/2}$, which gives $C(A,A) = 1$;
\item padding the waveforms with zero values to make them of identical length.
\end{enumerate}

\begin{figure}[t]
\centering
\includegraphics[height=0.37\linewidth]{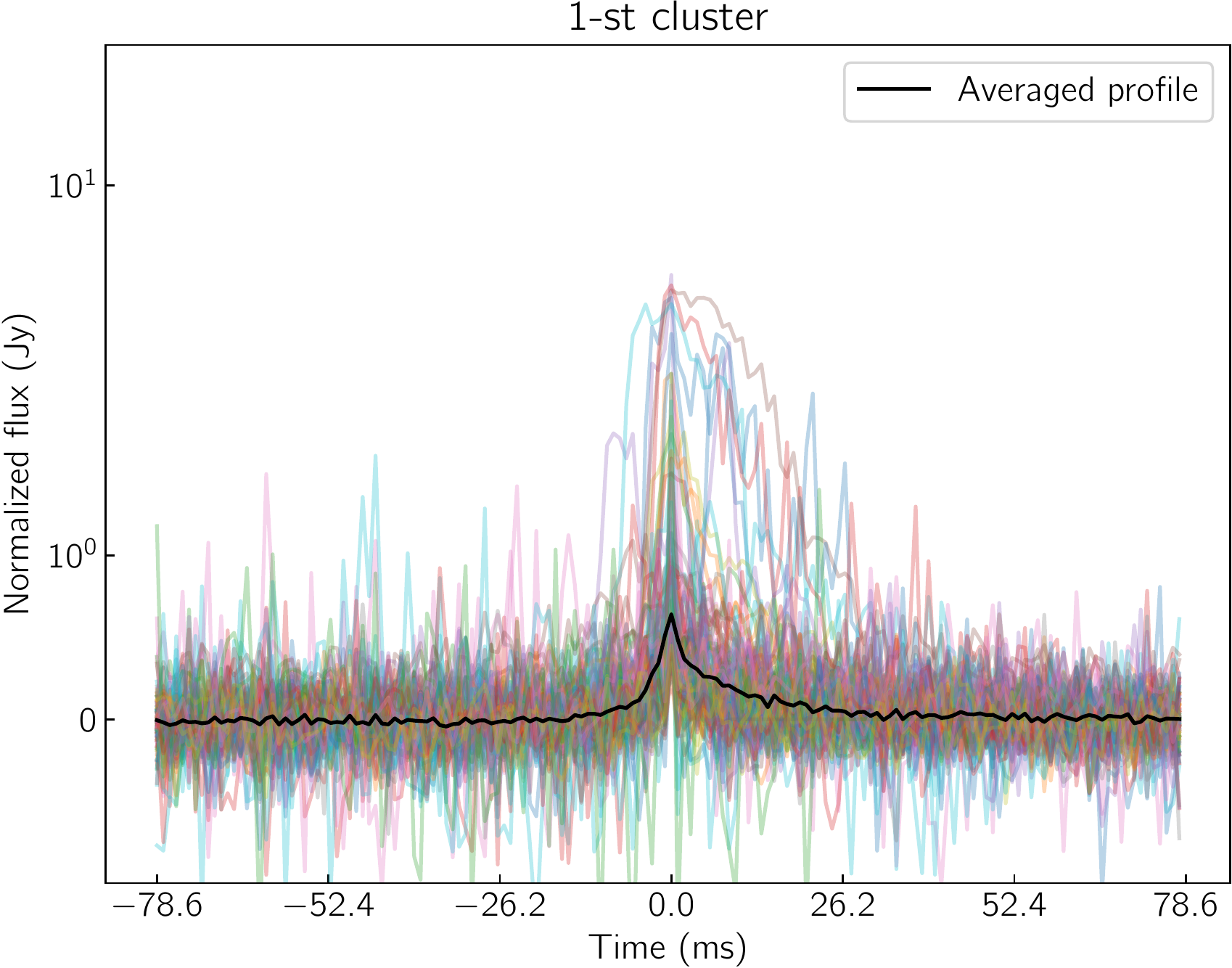}~~~\includegraphics[height=0.37\linewidth]{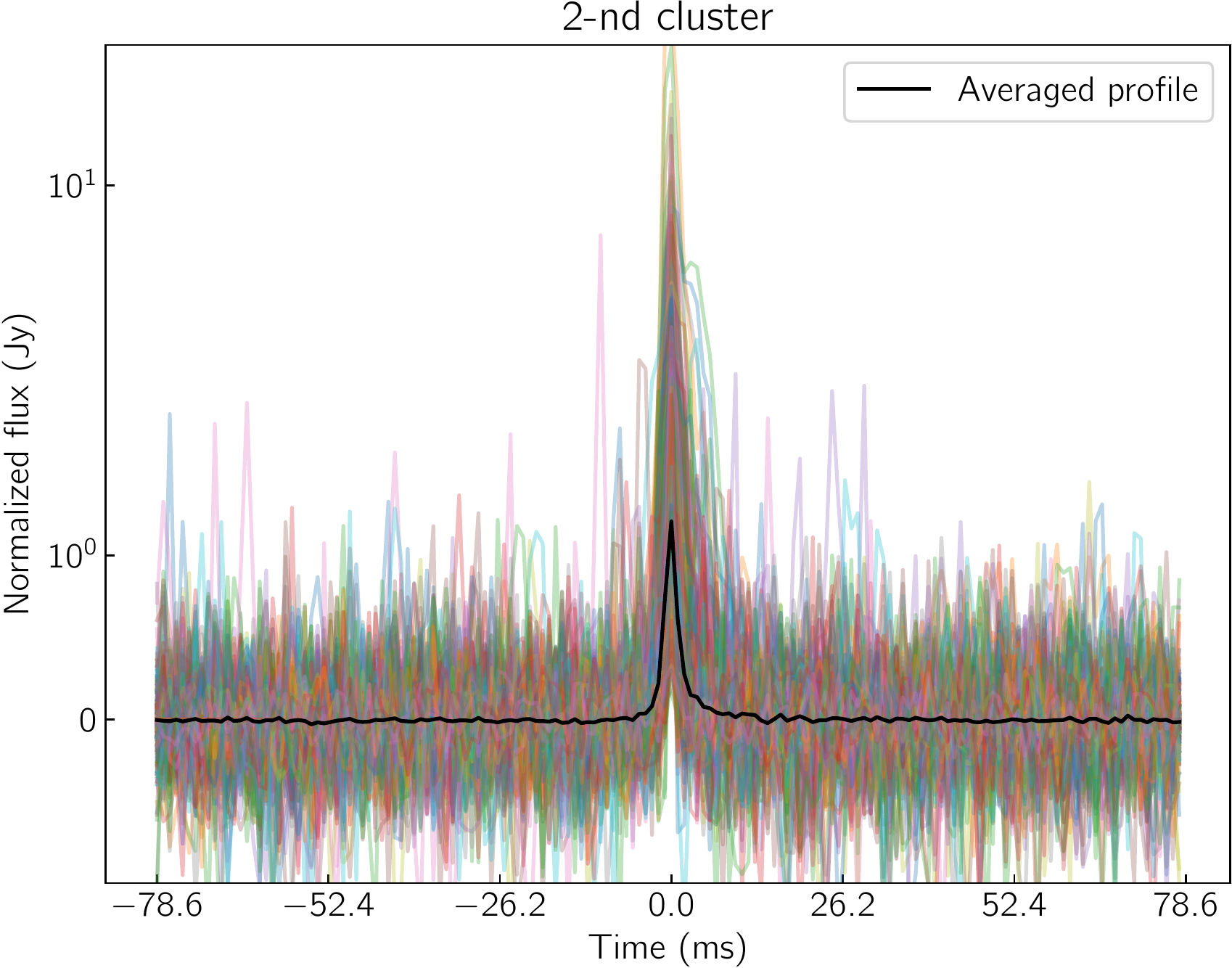}\\
~\\
\includegraphics[height=0.29\linewidth]{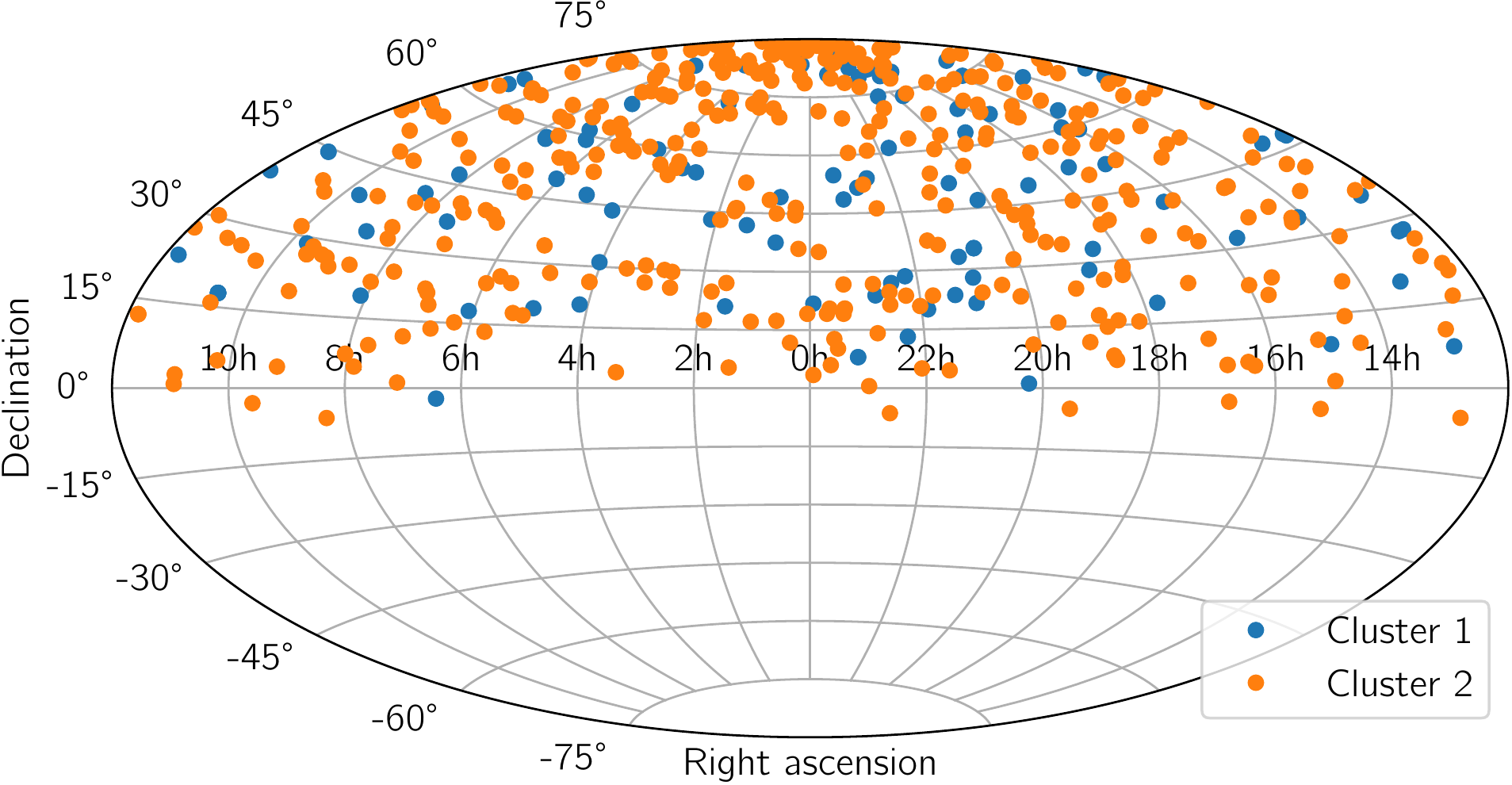}~~~\includegraphics[height=0.29\linewidth]{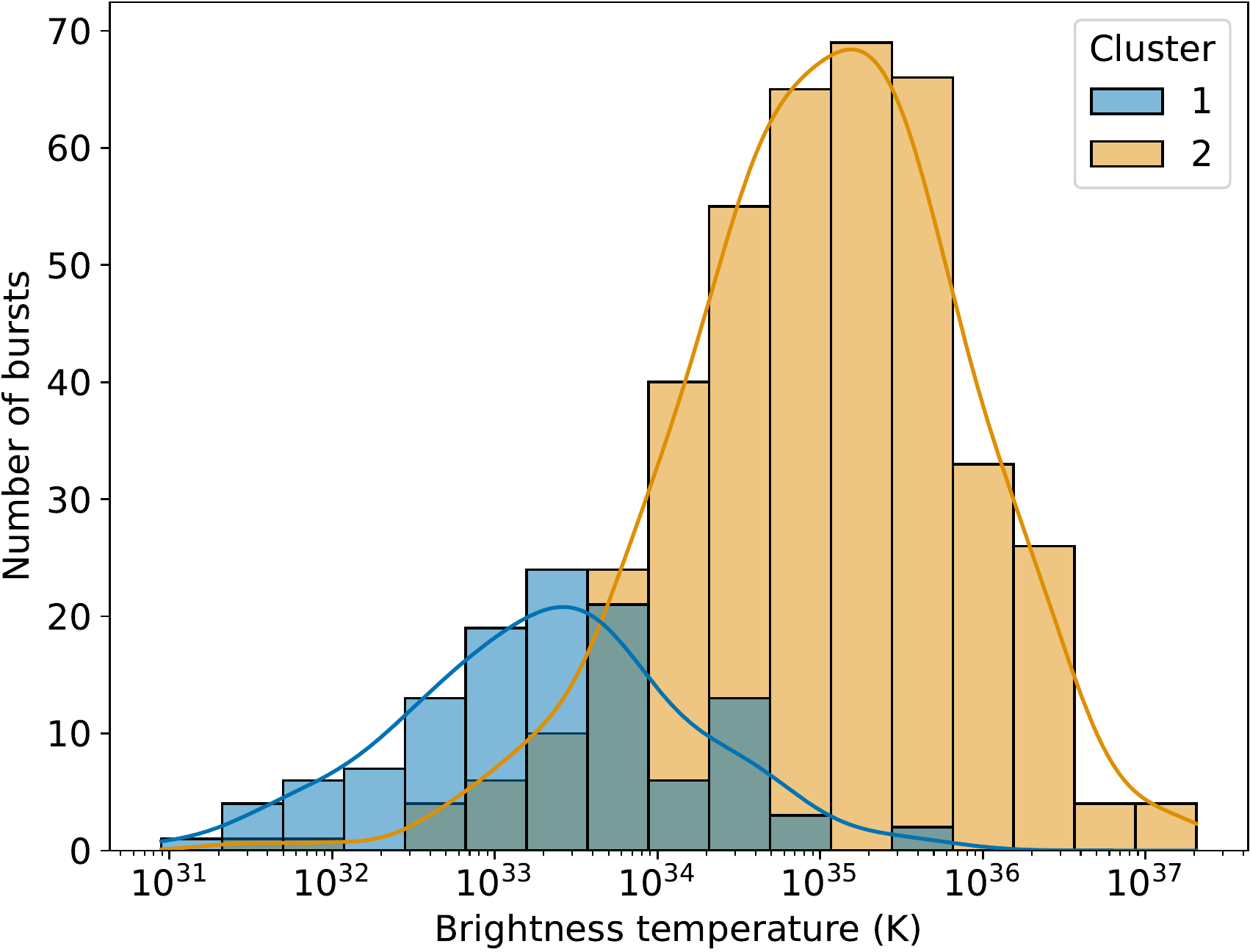}\\
\caption{Results of the two-cluster classification. \textit{Top:} Populations of  events in different clusters after baseline removal and normalization. The black solid lines in both top panels depict the average of the underlying profiles. \textit{Bottom left:} Localization of the classified events on the skymap. \textit{Bottom right:} Reconstructed brightness temperature distribution.}
\label{fig:results}
\end{figure}

The cross-correlation of each pair of FRB waveforms (including those from the repeaters) form the similarity matrix ${C_{ij} = C_{ji}}$, $C_{ii} = 1$, which is an input for the hierarchical clustering implemented with the Ward's method for the linkage \citep{doi:10.1080/01621459.1963.10500845}. The basis of this method is a reduction of the sum of squared distances between each observation and the average observation in a cluster.
The clusters are formed with a bottom-up approach, meaning that each of the observations starts as its own cluster, and clusters are merged together until meeting the threshold criteria according to the chosen linkage criteria, which determines the metric used for the merge strategy. 
The preprocessing and clustering are implemented in a corresponding software, \texttt{frbmclust}~\citep{frbmclust}.
For the present work we set threshold in such a way that allows for obtaining two major clusters (see~Fig.~\ref{fig:results}). This situation reveals significant differences in morphology visible even by unaided eye (see top panel of Fig.~\ref{fig:results}):
\begin{itemize}
\item The first cluster (119 events) is generally characterized by broader widths and lower fluxes, often featuring several peaks per event (13.4\% of the events have more than one peak). The mean boxcar width equals 24.79\,ms, the median flux --- 0.56\,Jy, number of repeaters in the cluster equals 28, five of which belong to FRB20180916B and seven --- to FRB20180814A;
\item The second cluster (408 events) is generally characterized by narrower widths and higher fluxes featuring mostly single peak per event (6.3\% of the events have more than one peak). The mean boxcar width equals 4.12 ms, the median flux --- 1.08 Jy, number of repeaters in the cluster equals 33, fourteen of which belong to FRB20180916B and three --- to FRB20180814A.
\end{itemize}

To test for possible systematic or instrument response biases which are possibly not taken into account (e.g., zenith-azimuth dependence), we produced a distribution of the clustered events in the sky, which seems to be isotropic (see the bottom left panel in Fig.~\ref{fig:results}).

A brightness temperature can be written as~\citet{2022A&A...657L...7X}:
\begin{equation}
    T_\mathrm{B} = 1.1 \times 10^{35} \mathrm{K} \left(\frac{F_{\nu}}{\mathrm{Jy}}\right) \left(\frac{T}{\mathrm{ms}}\right)^{-2} \left(\frac{\nu}{\mathrm{GHz}}\right)^{-2} \left(\frac{d_\mathrm{A}}{\mathrm{Gpc}}\right)^{2} \left(1+z\right)^{2},
\end{equation}
where $F_{\nu}$ is a flux density, $T$ is a boxcar width, thus $\nu$ is the emission frequency set to the constant value of 0.6 GHz (the central frequency of CHIME/FRB). The angular  distance $d_\mathrm{A}$ is derived from the redshift $z$ using~\texttt{FRB}~\citep{j_xavier_prochaska_2019_3403651} and $z$ is derived from the dispersion measure using \texttt{cosmocalc}~\citep{Hunter:2007}. The bottom right panel in Fig.~\ref{fig:results} shows the obtained distributions of $T_\mathrm{B}$.

\section{Discussion and conclusion}

In several papers the authors already studied statistical properties of FRBs demonstrating dichotomy among them using different sources and/or criteria. 
\cite{2019ApJ...885L..24C} 
and \cite{2020ApJ...891L...6F} 
discovered a statistical  differences in duration of one-off and repeating sources.
Recently, 
\cite{2022arXiv220204422Z} 
used a more complex approach to confront properties of the repeaters and non-repeaters. 
Some authors focused on particular sources with numerous detected bursts, especially on FRB 121102.
\cite{2021Natur.598..267L} 
found the energy distribution of radio flares from this source to be bimodal.
\cite{2022A&A...657L...7X} 
suggested that events from FRB 121102 with the brightness temperature above $\sim10^{33}$~K have different properties in comparison to those with $T_\mathrm{B}\lesssim 10^{33}$~K.

In this note we present a different result, not repeating any of those cited above. Both groups of the sources identified in our study include the repeating and non-repeating sources. Analysis of the repeaters demonstrated that they can produce events belonging to both of the identified clusters. 

However, the fact that events from the first cluster have smaller $T_\mathrm{B}$ and wider pulses resembles findings by \cite{2022A&A...657L...7X} (see, e.g., their Fig.~3), but for a broad population of sources. 
On the one hand, larger $T_\mathrm{B}$ for stronger shorter bursts is obvious from Eq.~1. On the other hand, the bimodal structure
of the $T_\mathrm{B}$ distribution in Fig.~1 might indicate physical difference between the bursts from the different clusters. 
It seems, that our cluster 2 corresponds to ``classical bursts'' in \cite{2022A&A...657L...7X}, and critical $T_\mathrm{B}$ as proposed by \cite{2022A&A...657L...7X} is different for different sources, and it covers the range $\sim10^{33}$--$10^{35}$~K. In our approach, two types of FRBs are identified without information about the brightness temperature, so, the critical value for each source is not fixed by hand. Thus, the division of bursts in two groups with respect to $T_\mathrm{B}$ appears without any {\it a priori} assumptions.
A more detailed analysis of statistical properties of the two clusters (and the possibility of existence of additional group of FRBs) will be presented elsewhere.

It is worthwhile to note that the signal cross-correlation used in our work is not optimized for the data provided in the catalog. 
Since we correlate the intensities, the phase information cannot be used, thus, the best correlation is achieved when several peaks overlap. 
Absence of the phase information also complicates the detection of weak signals in the presence of noise.
In future studies  we plan to test different approaches to mitigate these features of the data, e.g., by compression of a high signal-to-noise ratio peaks (similar to how it is done in audio engineering).

\begin{acknowledgments}
The authors thank V.~Lenok and M.~Pshirkov for the fruitful discussions.
\end{acknowledgments}

%

\vspace{5mm}
\facilities{CHIME(FRB)}


\software{astropy \citep{2013A&A...558A..33A,2018AJ....156..123A},  
          Cloudy \citep{2013RMxAA..49..137F}, 
          Source Extractor \citep{1996A&AS..117..393B}, SciPy \citep{2020SciPy-NMeth}, NumPy \citep{harris2020array}, Seaborn \citep{Waskom2021}, Matplotlib \citep{Hunter:2007}, cosmocalc \citep{2006PASP..118.1711W}, FRB \citep{j_xavier_prochaska_2019_3403651}, frbmclust \citep{frbmclust}}





\bibliography{sample631}{}
\bibliographystyle{aasjournal}



\end{document}